\documentclass[aps,prb,twocolumn,groupedaddress]{revtex4}

\usepackage{graphics}
\usepackage{graphicx}

\begin{document}

\title{ Tunable "Doniach Phase Diagram" for strongly-correlated nanoclusters}
\author{Yan Luo, Claudio Versozzi, and Nicholas Kioussis}
\email[E-mail me at: ]{yan.luo@csun.edu. }
\affiliation{Department
of Physics, California State University, Northridge, California
91330-8268}

\begin{abstract}
 Exact diagonalization calculations
reveal that the energy spacing $\Delta$ in the conduction band
tunes the interplay between the {\it local} Kondo and {\it non
local} RKKY interactions, giving rise to a ``Doniach phase
diagram" for a
 nanocluster with regions of prevailing Kondo
or RKKY correlations. The parity of the total number of electrons
alters the competition between the Kondo and RKKY correlations.
This interplay may be relevant to experimental realizations of
small rings or quantum dots with tunable magnetic properties.
Below a critical value V$_c$ of the hybridization the
susceptibility exhibits a low-T exponential activation behavior
determined by the interplay of the spin gap and $\Delta$.
\end{abstract}
\maketitle

The interplay between the {\it local} Kondo interactions and the
{\it non local} Ruderman-Kittel-Kasuya-Yosida (RKKY) interactions
is a central unresolved problem in the physics of dense valence
fluctuation and heavy fermion compounds.\cite{doniach,varma,si}
The first interaction, responsible for the quenching of the local
$f$-moment via the screening of the conduction electrons, has a
characteristic energy scale given by the single-impurity bulk
Kondo temperature, $T_K \propto exp(-1/\rho(E_F)J)$.\cite{Hewson}
Here, $\rho(E_F)$ is the density of states of the conduction
electrons at the Fermi energy and $J$ is the antiferromagnetic
interaction between the impurity spin and the conduction
electrons. The latter interaction
 is an indirect magnetic interaction between
localized moments mediated by the conduction electrons, with an
energy scale of order $T_{RKKY} \propto J^2\rho(E_F)$ which
promotes long- or short-range magnetic
ordering.\cite{doniach,varma,Hewson} In the high-temperature
regime the localized moments and the conduction electrons retain
their identities and interact weakly with each other. At
low-temperatures, the moments order magnetically if the RKKY
interaction is much larger than the Kondo energy, while in the
reverse case the system forms a heavy Fermi liquid of
quasiparticles which are composites of local moment spins bound to
the conduction electrons.\cite{si} The overall physics can be
described by the well-known ``Doniach phase
diagram".\cite{doniach} The description of the low-temperature
state, when both the RKKY and the Kondo interactions are of
comparable magnitude, is an intriguing question that remains
poorly understood and is the subject of current active
research.\cite{si}

There has been a revival of the Kondo physics in recent years due
to an exciting series of experiments on nanoscale systems, which
has enabled to probe the local interactions in a well controlled
way at the nanoscale.\cite{gordon,madhavan,manoharan,odom} For
example, STM experiments\cite{odom} studied the interaction of
magnetic impurities with the electrons of a single-wall nanotube
confined in one dimension. Interestingly, in addition to the bulk
Kondo resonance new sub peaks were found in {\it shortened} carbon
nanotubes, separated by about the average energy spacing,
$\Delta$, in the nanotube. These experiments invite an interesting
and important question: How is the interplay between the Kondo
effect and RKKY interactions becomes modified in a nanocluster
containing magnetic impurities, where the conduction electron
spectrum becomes discrete with a mean energy level spacing
$\Delta$. For such small systems, controlling $T_K$ upon varying
$\Delta$ is acquiring increasing importance since it allows to
tune the cluster magnetic behavior and to encode quantum
information. While the effect of $\Delta$ on the single-impurity
Anderson or Kondo model has received considerable
theoretical\cite{thimm, Hu, Balseiro, Affleck, Schlottman,
halperin}attention recently, its role on {\it dense} impurity
clusters remains an unexplored area thus far. The low-temperature
behavior of a nanosized heavy-electron system was recently studied
within the mean-field approximation\cite{schlottmann2001}.

%%%%%%%%%%%%%%%%%%%%%%%%%%%%%%%%%%%%%%%%%%%%%%%%%%%%%%%%%%%%%%
In this work we present exact diagonalization calculations for
$d$- or $f$-electron nanoclusters to study the effect of energy
spacing, parity of number of electrons, and hybridization on the
interplay between the Kondo and RKKY interactions in dense
strongly correlated electron nanoclusters. While the properties of
the system depend on their geometry and size\cite{Pastor}, the
present calculations treat exactly the Kondo and RKKY
interactions. Our results show that: i) tuning $\Delta$ and the
parity of the total number of electrons can drive the nanocluster
from the Kondo to the RKKY regime, i.e. a zero- temperature energy
spacing versus hybridization phase diagram; ii) the temperature
versus hybridization ``Doniach" phase diagram for the nanocluster
depends on the energy spacing ; and iii) Below a critical value
V$_c$ of the hybridization the susceptibility exhibits a low-T
exponential activation behavior determined by the interplay of the
spin gap and $\Delta$. In contrast, above $V_c$ there is no
exponential behavior.

The one dimensional Anderson lattice Hamiltonian is

\begin{eqnarray}
H =-t\sum_{i\sigma }(c_{i\sigma }^{\dagger }c_{i+1\sigma }+H.c)+
E_{f}\sum_{i\sigma }n_{i\sigma }^{\mathit{f}} \nonumber \\
+U\sum_{i}n_{i\uparrow }^{\mathit{f}}n_{i\downarrow }^{\mathit{f}%
} +V\sum_{i\sigma }(\mathit{f}_{i\sigma }^{\dagger }c_{i\sigma
}+H.c.),
\end{eqnarray}

\noindent where, t is the nearest-neighbor hopping matrix element
for the conduction electrons ($ \epsilon _{k}=-2tcosk$),
$f_{i,\sigma }$ $(c_{i,\sigma })$ annihilates a localized $f$
(conduction) electron on site i with spin $\sigma $, $E_{f}$ is
the energy level of the localized orbital, $V$ is the on-site
$f-c$ hybridization matrix element and $U$ is the on-site Coulomb
repulsion of the f electrons. We consider the half-filled ($N_{el}
= 2N = 12$) symmetric ($E_{f}=-\frac{U}{2}= -2.5$) case. As
expected, the cluster has a singlet ground state ($S_g=0$ where
$S_g$ is the ground-state spin) at half filling. The energy
spacing in the conduction band is $\Delta = 4t/(N-1) =
\frac{4t}{5}$. The exact diagonalization calculations employ
periodic boundary conditions. We compare the onsite {\it local}
Kondo spin correlation function (SCF) $<S_{i,z}^fS_{i,z}^c>$ and
the {\it nonlocal} (nearest-neighbor) RKKY SCF
$<S_{i,z}^fS_{(i+1),z}^f>$ to assign a state to the Kondo or RKKY
regimes, in analogy with mean field treatments.\cite{lacroix}

\indent In Fig. 1 we present the variation of the local Kondo SCF
$<S_{i,z}^fS_{i,z}^c>$ (squares) and the nearest-neighbor RKKY SCF
$<S_{i,z}^fS_{(i+1),z}^f>$ (circles) as a function of
hybridization for two values of the hopping matrix element $t=0.2$
(closed symbols) and $t=1.2$ (open symbols), respectively. For
weak hybridization the nonlocal (local) nearest-neighbor RKKY
(Kondo) SCF is large (small), indicating strong short-range
antiferromagnetic coupling between the
 $f-f$ local moments, which leads to long range magnetic ordering for
 extended systems. As V increases,
$<S_{i,z}^fS_{(i+1),z}^f>$ decreases whereas the
$<S_{i,z}^fS_{i,z}^c>$ increases (in absolute value) saturating at
large values of V. This gives rise to the condensation of
independent local Kondo singlets at low temperatures, i.e., a
disordered spin liquid phase. For large $V$ the physics are {\it
local}. Interestingly, as $t$ or $\Delta$ decreases the {\it f-c}
SCF is dramatically enhanced while the {\it f-f} SCF becomes
weaker, indicating a transition from the RKKY to the Kondo regime.

\begin{figure}[ht]
\centerline{\includegraphics[width=.55\textwidth]{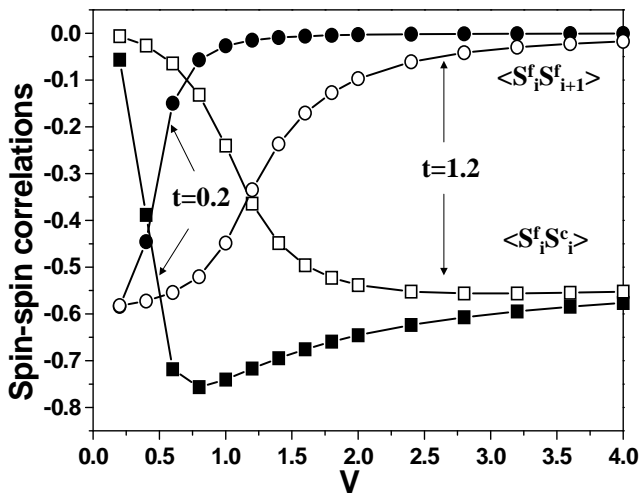}}
\vspace {-0.3 in} \caption{Nearest-neighbor f-f SCF (circles) and
on-site f-c SCF (squares) as a function of V for $t=0.2$ (closed
symbols) and $t=1.2$ (open symbols), respectively.} \label{fig1}
\end{figure}

In Fig. 2 we present the energy spacing versus V zero-temperature
phase diagram of the nanocluster, which illustrates the interplay
between Kondo and RKKY interactions. In the RKKY region
$<S_{i,z}^fS_{(i+1),z}^f>$ is larger than  $<S_{i,z}^fS_{i,z}^c>$
and the {\it total} local moment squared $\mu^2 = \langle
(\mu_{f}+\mu_{c})^{2}\rangle \ne 0$; in the Kondo regime
$<S_{i,z}^fS_{(i+1),z}^f>$ is smaller than the
$<S_{i,z}^fS_{i,z}^c>$, $\mu^2=0$, and the ground state is
composed of independent local singlets. The solid curve indicates
the critical value of the hybridization $V=V_c$ or the critical
value of energy spacing $\Delta=\Delta_c$, where the {\it local}
and {\it non local} SCF's are equal, i.e.,
$<S_{i,z}^fS_{(i+1),z}^f> = <S_{i,z}^fS_{i,z}^c>$. The dashed
curve denotes the points where the {\it total} local moment square
vanishes. Thus, in the intermediate regime, the so-called {\it
free spins} regime \cite{schroder}, $<S_{i,z}^fS_{(i+1),z}^f>$ is
smaller than the $<S_{i,z}^fS_{i,z}^c>$, the $f$ moment is {\it
partially} quenched and $\mu^2 \not= 0$. Interestingly, we find
that
 the {\it free spins} regime becomes narrower as
the average level spacing $\Delta$ is reduced. This result may be
interpreted as a quantum critical regime for the nanocluster due
to the finite energy spacing, which eventually reduces to a
quantum critical point when $\Delta \rightarrow 0$.

\begin{figure}[ht]
\centerline{\includegraphics[width=0.55 \textwidth]{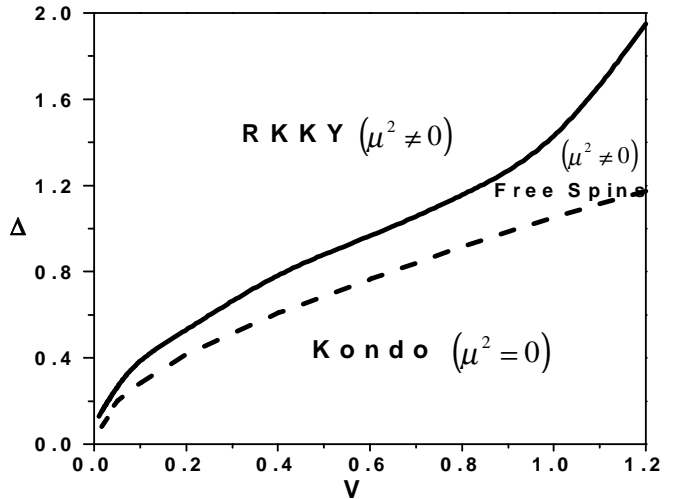}}
\vspace {-0.4 in}
 \caption{Energy spacing $\Delta$
versus hybridization zero-temperature phase diagram. The solid
curve denotes the crossover point of the spin-spin correlation
function in Fig.1; the dashed curve denotes the set of points
where the on-site {\it total} moment square $\langle
(\mu_{f}+\mu_{c})^{2} \rangle = 0.0 \pm 0.05$.} \label{fig2}
\end{figure}

The T=0 exact diagonalization results on small clusters are
generally plagued by strong finite size
effects\cite{Pastor,haule}. Performing calculations at finite T
gives not only the thermodynamic properties of the system, but
most importantly diminishes finite-size effects for $k_{B}T \gg
\Delta$. In Fig. 3 we show the nearest-neighbor f-f SCF and
on-site Kondo f-c SCF as a function of temperature for $t=0.2$ and
$V= 0.2 < V_c$ and $V= 0.4
> V_c$, where $V_c=0.25$. At high temperatures, the free
moments of the f and conduction electrons are essentially
decoupled. The nearest-neighbor {\it non local} SCF falls more
rapidly with $T$ than the on-site {\it local} $f-c$ SCF,
indicating that the {\it non local} spin correlations are
destroyed easier by thermal fluctuations. For $V<V_c$, the
nanocluster is dominated by RKKY interactions at temperatures
lower than the crossover temperature, $T^{cl}_{RKKY}$, which
denotes the temperature where the {\it non local} and {\it local}
interaction become equal in the nanocluster. In the infinite
system this temperature would denote the ordering N$\acute{e}$el
temperature. On the other hand, for $V>V_c$ the RKKY and Kondo
spin correlation functions do not intersect at any $T$, and the
physics are dominated by the local Kondo interactions.
\begin{figure}[ht]
\centerline{\includegraphics[width=.55 \textwidth]{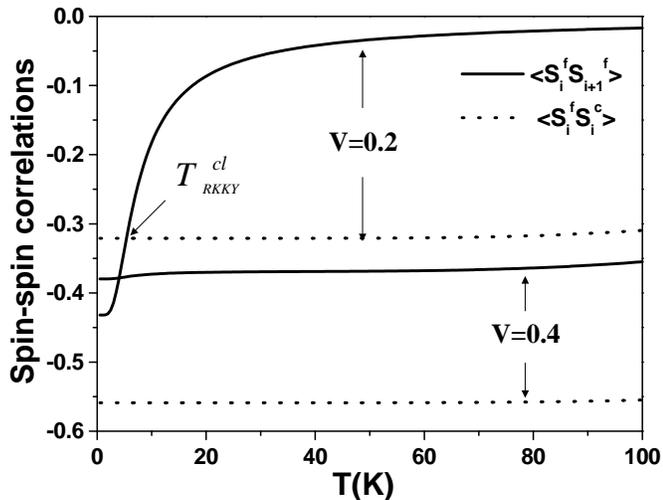}}
\vspace {-0.4 in}
 \caption{Nearest-neighbor f-f and on-site f-c SCF versus temperature for
$t=0.2$ and $V=0.2<V_c$ and $V=0.4>V_c$, $V_c=0.25$.} \label{fig3}
\end{figure}

%%%%%%%%%%%%%%%%%%%%%%%%%%%%%%%%%%%%%%%%%%%%%%%%%%%%%%%%%%%%

 In Fig. 4 we present the crossover temperature $T^{cl}_{RKKY}$ for the
cluster as a function of hybridization for different values of
$t$. This represents the phase diagram of the strongly correlated
nanocluster, which is similar to the ``Doniach phase diagram" for
the infinite Kondo necklace model.\cite{doniach} The phase for
$T<T^{cl}_{RKKY}$ denotes the regime where the {\it non local}
short-range magnetic correlations are dominant. For $V<V_c$ and
$T>>T^{cl}_{RKKY}$ one enters into the disordered ``free" local
moment regime. On the other hand, for $V>V_c$ and at low $T$, the
 nanocluster can be viewed as a condensate of singlets,
typical of the Kondo spin-liquid regime.  Interestingly, the
$T^{cl}_{RKKY}$ can be tuned by the energy spacing $\Delta$ or the
size of the cluster. Thus, increasing $\Delta$ or decreasing the
size of the nanocluster results to enhancement of the {\it non
local} nearest-neighbor RKKY correlations and hence
$T^{cl}_{RKKY}$. This result is the first exact ``Doniach phase
diagram" for a nanocluster.

\begin{figure}[ht]
\centerline{\includegraphics[width=.55\textwidth]{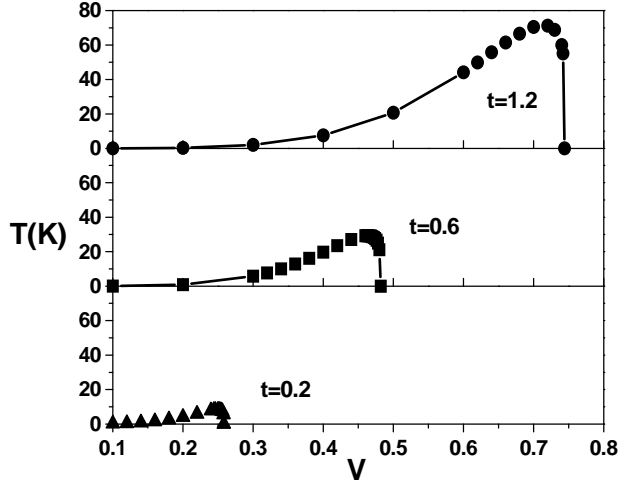}}
\vspace {-0.3in}
 \caption{Effect of energy spacing, $\Delta=\frac{4t}{N-1}$
on the exact ``Doniach phase diagram" for a strongly correlated
electron nanochain. The curve represents the  crossover
temperature $T^{cl}_{RKKY}$, where the {\it non local} short range
AF spin correlations become equal to the {\it local} on-site Kondo
spin correlations.} \label{fig4}
\end{figure}
%%%%%%%%%%%%%%%%%%%%%%%%%%%%%%%%%%%%%%%%%%%%%%%%%%%%%%%%%%%%
%%%%%%%%%%%%%%%%%%%%%%%%  Susceptibility and specific heat
%%%%%%%%%%%%%%%%%%%%%%%%%%%%%%%%%%%%%%%%%%%%%%%%%%%%%%%%%%%%%

In bulk Kondo insulators and heavy-fermion systems, the low-$T$
 susceptibility  and specific heat behavior is determined by the spin
 gap, which for the half-filled Anderson lattice model,
 is determined by the ratio of $V$ to $U$.
 On the other hand, strongly correlated nanoclusters
 are inherently associated with a new low-energy cutoff, namely
the energy spacing $\Delta$ of the conduction electrons. Thus, a
key question is how can the low-temperature physics be tuned by
the interplay of the spin gap and the energy spacing. The
temperature-dependent local f-spin susceptibility, $\chi^f(T)$,is
\begin{eqnarray}
\frac{ k_BT\chi^f(T )}{(g\mu_B)^2}= \frac{1}{Q} \sum_{\alpha}
e^{-\frac{E_{\alpha}}{k_BT}}<\alpha|S^f(i)S^{Tot}|\alpha>,
\end{eqnarray}
where $Q$ is the partition function. Here, $S^{Tot}$ is the
z-projection of the total spin (both the $f$- and
$c$-contributions), and $|\alpha>$ and $E_{\alpha}$ are the exact
many-body eigenstates and eigenvalues, respectively. When $V=0$,
the localized spins and conduction electrons are decoupled and
$\chi^f(T)$ is simply the sum of the Curie term due to the free f
spins and the Pauli term of the free conduction electrons. For
finite $V$, $\chi^f(T)$ decreases with temperature at
low-temperatures. In Fig. 5 we present $\chi^f(T)$  as a function
of temperature for $t=0.2$ and $V= 0.2 < V_c$, $V=V_c=0.25$, and
$V=0.4 > V_c$. For small $V$ the spin gap which is smaller than
$\Delta$ controls the exponential activation behavior of $\chi^f$
at low $T$ (see inset in Fig. 5). In contrast, in the large $V$
limit, the spin gap becomes larger than $\Delta$ and the low-$T$
behavior of the susceptibility shows no exponential activation. At
high $T$ we can see an asymptotic Curie-Weiss regime, typical of
localized decoupled moments. The low-$T$ behavior of the
susceptibility is associated with the lowest energy scale
determined by the lowest value between the spin gap and the energy
spacing $\Delta$. For large values of $t$ (or $\Delta$) the spin
gap is reduced and the spin gap is the lowest energy scale.
Consequently, the low-$T$ behavior exhibits exponential activation
associated with the spin gap. Similar low-T behavior is also found
for the specific heat.\cite{luo}

\begin{figure}[ht]
\centerline{\includegraphics[width=.55\textwidth]{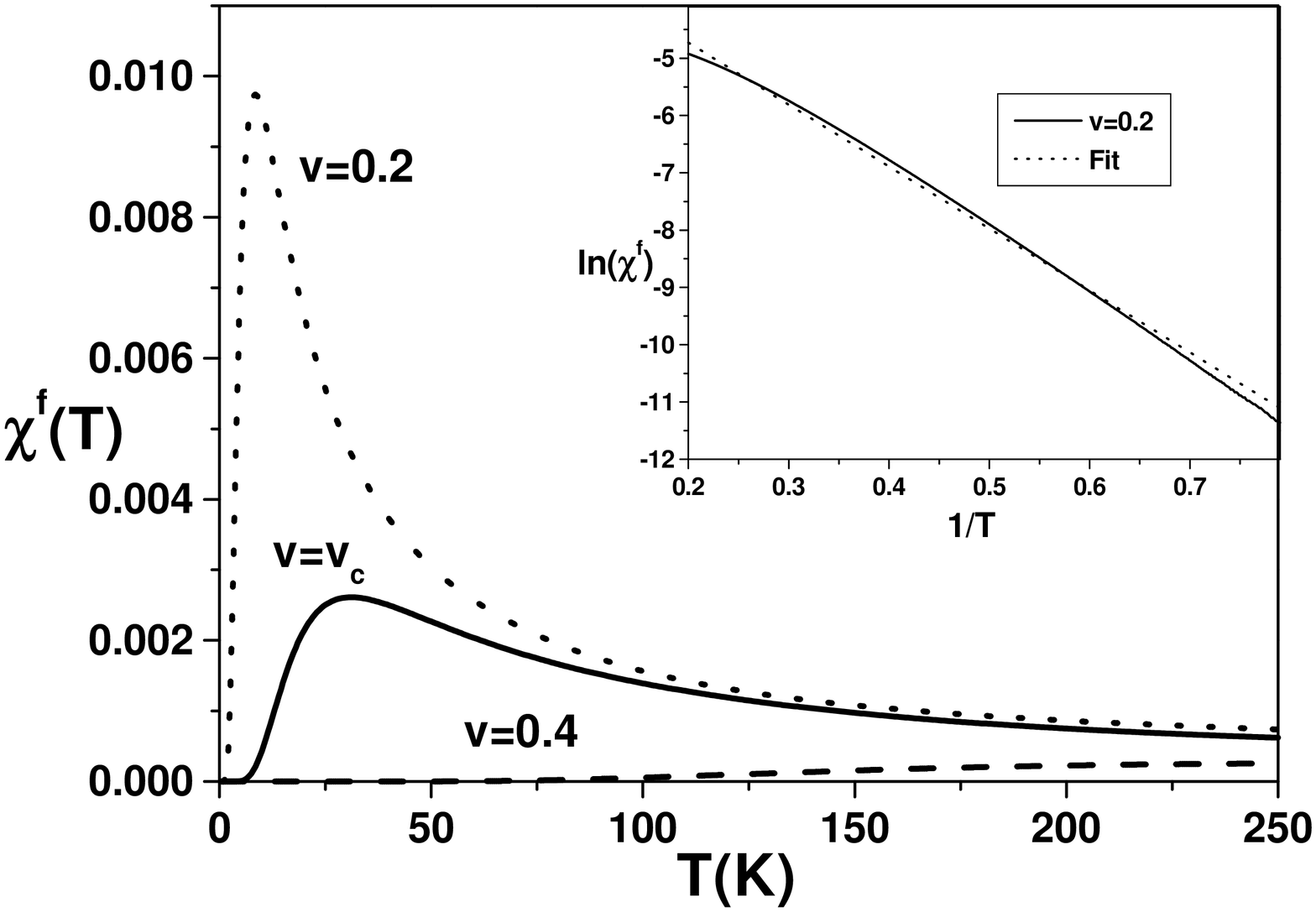}}
\vspace{-0.45 in}
 \caption{Local f magnetic susceptibility as a
function of temperature for $t=0.2$ and $V=0.2 < V_c$, $V = V_c =
0.25$ and $V=0.4>V_c$. The inset shows the linear fit for
$ln(\chi^{f})$ versus $\frac{1}{T}$ at the low temperature.}
\label{fig5}
\end{figure}

We have also examined the role of even versus odd number of
electrons on the magnetic behavior. For $t=1$, changing the number
of electrons from N$_{el}$ = 12 to N$_{el}$ = 11 results in: (a)
an enhancement of the local Kondo f-c spin correlation function;
and (b) a suppression of the nearest-neighbor f-f spin correlation
function. This interesting novel tuning of the magnetic behavior
can be understood as follows:  For an odd-electron cluster, the
topmost occupied single particle energy level is singly occupied.
On the other hand, for an even-electron cluster, the topmost
occupied single-particle energy level is doubly occupied, thus
blocking energy-lowering spin-flip transitions.  This energy
penalty intrinsically weakens the Kondo correlations\cite{thimm}.
As expected, changing the number of electrons from even to odd
changes $S_g$ = 0 to $S_g= \frac{1}{2}$. For t= 0.05 (Kondo
regime), the on site f-c correlation function does not depend as
strongly on the parity in the number of electrons because the
sites are locked into singlets. On the other hand, the {\it
nonlocal} RKKY becomes suppressed as in the case of large energy
spacing.

Exact diagonalization calculations reveal that the energy spacing
and the parity of the number of electrons, give rise to a novel
tuning of the magnetic behavior of a {\it dense} strongly
correlated nanocluster.  This interesting and important tuning can
drive the nanocluster from the Kondo to the RKKY regime, i.e. a
tunable ``Doniach" phase diagram in nanoclusters. For weak
hybridization, where the spin gap is smaller than $\Delta$, both
the low-temperature local $f$ susceptibility and specific heat
exhibit an exponential activation behavior associated with the
spin gap. In contrast in the large hybridization limit, $\Delta$
is smaller than the spin gap, the physics become local and the
exponential activation behavior disappears.  We believe that the
conclusions of our calculations should be relevant to experimental
realizations\cite{odom} of small clusters and quantum dots.

The research at California State University Northridge was
supported through NSF under Grant Nos. DMR-0097187, NASA under
grant No. NCC5-513, and the Keck and Parsons Foundations grants.
The calculations were performed on the the CSUN Massively Parallel
Computer Platform supported through NSF under Grand No.
DMR-0011656.

\end{document}